\documentclass[a4paper]{article}

\usepackage{ISCSLP2022}
\usepackage{color}
\usepackage{hyperref}
\usepackage{stfloats}

\title{AdaVITS: Tiny VITS for Low Computing Resource Speaker Adaptation}
\name{Kun Song$^1{^,}{^3}$, Heyang Xue$^2$, Xinsheng Wang$^1$, Jian Cong$^1$, Yongmao Zhang$^1$, Lei Xie$^1{^,}{^2}{^*}$, Bing Yang$^3$, Xiong Zhang$^3$, Dan Su$^3$}
\address{
  Audio, Speech and Language Processing Group, $^1$School of Computer Science, $^2$School of Software, Northwestern Polytechnical University, Xi'an, China\\
  $^3$Cloud and Smart Industries Group, Tencent Technology Co., Ltd., China}
\email{kunsong.npu.se@gmail.com, lxie@nwpu.edu.cn}

\begin{document}

\maketitle
\begin{abstract}
Speaker adaptation in text-to-speech synthesis (TTS) is to finetune a pre-trained TTS model to adapt to new target speakers with limited data. While much effort has been conducted towards this task, seldom work has been performed for low computational resource scenarios due to the challenges raised by the requirement of the lightweight model and less computational complexity. In this paper, a tiny VITS-based~\cite{DBLP:conf/icml/KimKS21} TTS model, named AdaVITS, for low computing resource speaker adaptation is proposed. To effectively reduce the parameters and computational complexity of VITS, an inverse short-time Fourier transform (iSTFT)-based wave construction decoder is proposed to replace the upsampling-based decoder which is resource-consuming in the original VITS. Besides, NanoFlow is introduced to share the density estimate across flow blocks to reduce the parameters of the prior encoder. Furthermore, to reduce the computational complexity of the textual encoder, scaled-dot attention is replaced with linear attention. To deal with the instability caused by the simplified model, we use phonetic posteriorgram (PPG) as a frame-level linguistic feature for supervising the model process from phoneme to spectrum. Experiments show that AdaVITS can generate stable and natural speech in speaker adaptation with 8.97M model parameters and 0.72 GFlops computational complexity. \footnote{Audio samples are available at \url{https://AdaVITS.github.io/AdaVITS/}}
\end{abstract}
\noindent\textbf{Index Terms}: speaker adaptation, low computing resource, adversarial learning, normalizing flows
\renewcommand{\thefootnote}{\fnsymbol{footnote}}
\footnotetext{Lei Xie is the Corresponding author.}

\vspace{-5pt}
\section{Introduction}
\label{sec:intro}

To create human-like natural speech, modern neural network-based text-to-speech (TTS) models are usually large and contain mass of parameters~\cite{DBLP:conf/interspeech/WangSSWWJYXCBLA17, DBLP:conf/nips/RenRTQZZL19, DBLP:conf/nips/KumarKBGTSBBC19, DBLP:conf/interspeech/SuJF20}. To train such a model, sufficient data and computational resources are necessary. However, to realize customized TTS, a corpus with sufficient samples recorded by a new target speaker is not always available in practice, which makes the few/one/zero-shot methods gain much interest. As compared with one/zero-shot methods~\cite{DBLP:conf/nips/ArikCPPZ18, DBLP:conf/iclr/ChenASBRZWCTLGO19} which usually rely on an extra speaker embedding module, \textit{speaker adaptation}, i.e, fine-tuning a well-trained base model with limited data to adapt to the new speaker, is still a practical approach with better speaker similarity. Considering the limited computational resources in many real-world scenarios, such as personalized voice services on edge devices, a lightweight TTS model with a small model size and low computation consumption is essential for the speaker adaptation task.


Due to the significant role of speaker adaptation in many scenarios, i.e., virtual avatars and personal assistants, much effort has been conducted in this field. For example,  AdaSpeech~\cite{DBLP:journals/corr/abs-2103-00993} reduces the number of adaptive parameters to alleviate the memory usage and serving cost by using conditional layer normalization. Besides, some approaches aim to reduce the model training time in the adaptation process to get a better user experience. For instance, Meta-Voice~\cite{DBLP:journals/corr/abs-2111-07218} uses meta-learning to obtain better model initialization for faster adaptation. Furthermore, some studies address the noise-robust speaker adaptation problem, aiming to obtain a noise-invariant TTS model for the target speaker with only noisy samples at hand~\cite{DBLP:conf/interspeech/CongY0YW20}.

While these efforts have successfully improved the performance of speaker adaptation models with limited data, reducing the TTS model size and computation complexity is still desired because the current popular TTS models are still too large. For instance, a typical Fastspeech 2~\cite{DBLP:conf/iclr/0006H0QZZL21} model has 28M parameters, while the use of a neural vocoder adds extra. It is non-trivial to obtain a lightweight solution for speaker adaptation with decent performance. First, effectively reducing the parameters and computational complexity is an intuitive challenge. Second, the simplified model structure could arise instability in the generation of speech, resulting in low speaker similarity and naturalness with obvious artifacts and even pronunciation errors. Model compression via distillation and quantization~\cite{DBLP:journals/corr/HintonVD15} can be directly adopted but it may induct apparent performance loss. Recent effort on neural architecture search (NAS)~\cite{DBLP:conf/icassp/Luo0WQLZCL21} is another promising solution while the search process itself consumes much computation power and time effort.

To face the above challenges, in this paper we propose a tiny TTS model for low computing resource speaker adaptation. Inspired by the superiority of VITS~\cite{DBLP:conf/icml/KimKS21}, which is a fully end-to-end TTS model, on eliminating the mismatch between acoustic feature generation and wave construction in typical two-stage based methods, our lightweight solution, named \textit{AdaVITS}, is built upon VITS with substantial modification to fit the speaker adaption scenario with fewer parameters, lower computational complexity, and stable performance.  First, considering the resource-consuming characteristic of the upsampling-based decoder, an inverse short-time Fourier transform (iSTFT)-based wave construction decoder is proposed. Besides, to reduce the parameters of the prior encoder, flow indication embedding (FLE) is utilized to share the density estimate across flow blocks. Moreover, for the FFT blocks, scaled-dot attention is replaced with linear attention to reduce computational complexity. To deal with the instability caused by the simplified model, phonetic posteriorgram (PPG) is used as a frame-level linguistics feature to constrain the phoneme to spectrum modeling process. Extensive experiments demonstrate the good performance of AdaVITS on the speaker adaptation task with only 8.97M model parameters and 0.72 GFlops computation.

\section{Method}
\label{sec:format}

\begin{figure}[ht]
\setlength{\abovecaptionskip}{0.cm}
	\centering
	\includegraphics[width=8cm,height=3.2cm]{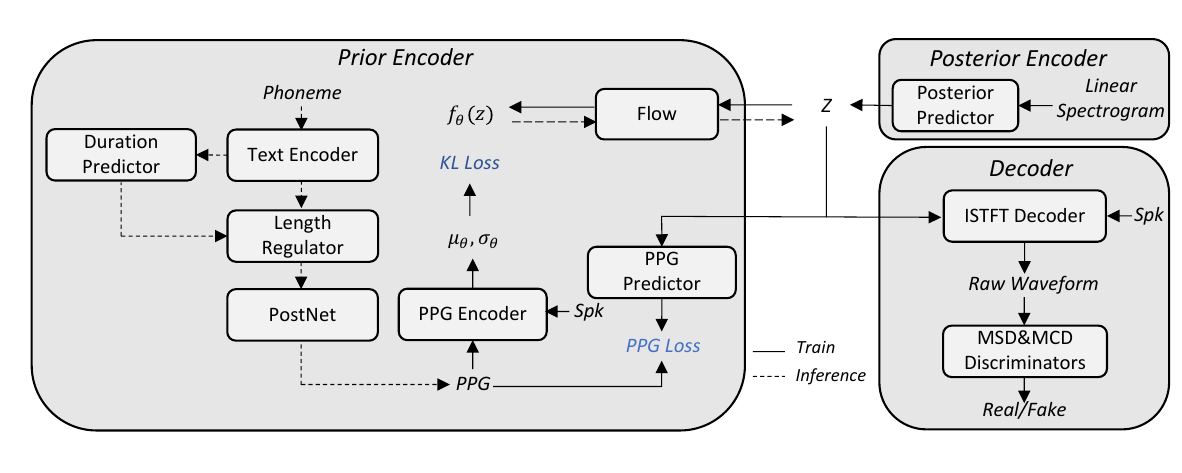}
	\caption{
		Architecture of AdaVITS.
		}
	\label{model_flatts}
\end{figure}

VITS~\cite{DBLP:conf/icml/KimKS21} is an end-to-end model of state-of-the-art, which uses variational autoencoder to learn latent variable as an intermediate representation between acoustic model and vocoder in end-to-end learning. In order to make the prior distribution close to the latent variable $z$, VITS added normalizing flow to the prior network to improve the representation ability of the prior distribution. Inspired by VITS, in this paper, we use the VITS approach to model the process from text features to speech waveforms, with the difference of using frame-level text features PPG as the intermediate constraint between phoneme and $z$. The advantage of using PPG is that it can explicitly decouple the timbre and content information of speech, which will make modeling more flexible~\cite{WangTFYWZ20}. In addition, in speaker adaptation, we can only learn the target speaker's timbre rather than specific speech characteristics, for many people do not have reasonable control of speed and prosody when recording. As illustrated in Figure~\ref{model_flatts}, similar to VITS, the proposed AdaVITS is composed of a posterior encoder, a prior encoder, and a decoder. The posterior encoder is used to extract latent variable $z$ from the waveform when training and is not used in inference. The prior encoder is used to extract the prior distribution $p(z|c)$ of $z$ from the phoneme, and the decoder is performed to generate waveform by the $z$ and speaker embedding. 
\vspace{-0.1cm}

\vspace{-8pt}
\subsection{Posterior Encoder}
\vspace{-3pt}
In AdaVITS, the posterior encoder is similar to that in the original VITS, which takes the linear spectrum as input to extract the mean and variance of the posterior distribution $p(z|y)$, and obtains the latent variable $z$. Due to the already existing  speaker information in the linear spectrum, no extra speaker embedding is attached.

\subsection{Prior Encoder}
Conditioned on the conditional information $c$, including phoneme and speaker embedding, the prior encoder is to get the prior distribution $p(z|c)$ of the conditional variational autoencoder (CVAE). Compared with the original VITS, we use PPG as an intermediate constraint from phoneme to $z$. PPG is extracted from the acoustic model of a speaker-independent automatic speech recognition system, which is a frame-level linguistic feature not contain speaker information. Compared with spectrums, which contain not only linguistic information but also rich acoustic information, less information conveyed by PPG allows us to greatly simplify the complexity of the model and decouple the content and speaker information.

For text processing, fewer FFT layers are used for the text encoder, and then a length regulator is used to expand features from phoneme level to frame level, the construction of PPG is directly performed by the post-net rather than the structure with decoder and post-net. Since the computational complexity of scaled-dot attention in FFT is not linear with the sequence length $n$, it has incredibly high computational complexity in long sentences. Here, referring to~\cite{DBLP:conf/icml/KatharopoulosV020}, linear attention is used to replace scaled-dot attention in FFT blocks, which will ensure the attention effect while reducing computational complexity. The modeling process from phoneme to PPG is pre-trained and not finetuned in speaker adaptation.

After obtaining PPG, the PPG encoder is used to get a prior normal distribution with mean $\mu_{\theta}$ and variance $\sigma_{\theta}$ from PPG and speaker embedding. To be specific, the PPG encoder is  composed of FFT blocks, in which linear attention is utilized here to reduce the computational complexity.

Due to the lack of explicit constraint of the pronunciation to $z$, the model tends to raise pronunciation errors such as mispronunciation and abnormal tone. To face this issue,  a PPG predictor is introduced to provide the pronunciation constraint. The architecture of the PPG predictor is consistent with the phoneme predictor of VISinger~\cite{DBLP:journals/corr/abs-2110-08813}. With the input of $z$, the $\hat{\text{PPG}}$ produced by the PPG predictor is used to obtain the following constraint loss:
\begin{equation}
  \setlength{\abovedisplayskip}{3pt}
  \setlength{\belowdisplayskip}{3pt}
  L_{\text{ppg}}=\left\| \hat{\text{PPG}} - \text{PPG} \right\|_{1}\label{LPPG}
\end{equation}
where PPG is the input. The PPG predictor is only trained on the pre-trained model and will be frozen during the adaptation. 

The distribution  $z$ is then transformed into a more complex distribution using the normalized flow $f_{\theta}$. This normalized flow includes multiple layers of affine coupling, and each layer consists of a stack of WaveNet~\cite{DBLP:conf/ssw/OordDZSVGKSK16} residual blocks following VITS. Due to the use of multiple layers, the flow has a large number of parameters. Referring to NanoFlow~\cite{DBLP:conf/nips/LeeKY20}, we share the parameter in each affine coupling layer of flows, and each layer is distinguished by FLE. The amount of parameters in flow is controlled to be a single layer through this method. 
Similar to VITS, latent variable $z$ is transformed to $f(z)$ by flow during training. During inference, the output of the PPG encoder is transformed into a latent variable $\hat{z}$ by the inverse flow. As the speaker embedding is added to the input PPG, no extra speaker embedding will be added to the flow.
\begin{equation}
  \setlength{\abovedisplayskip}{3pt}
  \setlength{\belowdisplayskip}{3pt}
  p(z|c) = N(f_\theta(z);\mu_\theta(c), \sigma_\theta(c))) \big \vert det \frac{\partial f_{\theta}(z)}{\partial_z } \big \vert .
\end{equation}
\vspace{-0.4cm}

\vspace{-3pt}
\subsection{Decoder}
\label{sec2:decoder}

In VITS, the reconstruction of waveform is performed by a typical vocoder-like decoder, which consists of a series of upsampling layers. While this upsampling layers-based decoder generally has strong modeling capabilities, the gradual increasing process to transfer the input to the time domain is computation consuming. Because our model is equivalent to the joint training of the acoustic model and the vocoder, it is feasible to use iSTFT to generate the waveform directly. In practice, the real and imaginary parts of the waveform are predicted based on the features in the frequency domain, which can effectively reduce the computational cost. 

\begin{figure}[ht]
\setlength{\abovecaptionskip}{0.1cm}
	\centering
	\includegraphics[width=7.245cm,height=5.67cm]{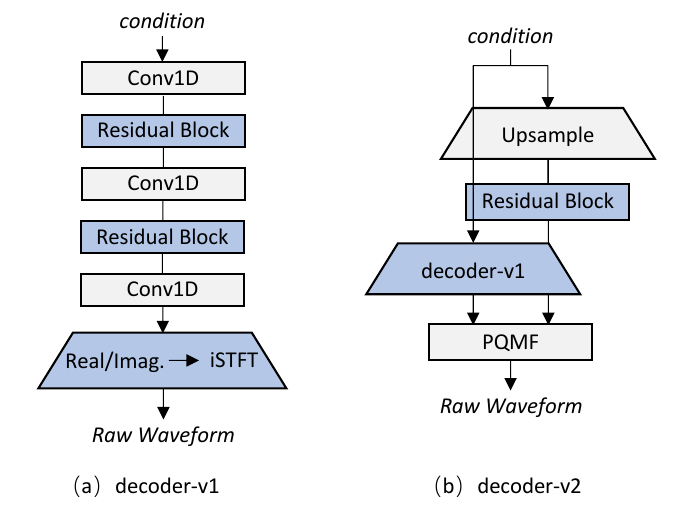}
	\caption{
		Architecture of decoder.
	}
	\label{decoder}
\end{figure}

As illustrated in Figure~\ref{decoder}(a), we propose decoder-v1. We use multiple convolutions to gradually increase the input dimension to $(f/2+1)*2$ to make the output fit the total dimension of real and imaginary parts, where $f$ indicates the fast Fourier transform size. A stack of residual blocks follow each one-dimensional convolution for more information on the corresponding scale. Due to the frequency domain dimension modeling, we do not use dilated convolution but use a smaller kernel size with the aim to ensure that the receptive field will not be too large. The group convolution is used in one-dimensional convolution to save computation. Then, the output is split into real and imaginary parts, based on which the final waveform can be produced via iSTFT. Note that, following VITS, the input condition includes speaker embedding and latent variables $z$, as we found that the speaker similarity will be degraded significantly if the speaker embedding is not added to the decoder.  

As illustrated in Figure~\ref{decoder}(b), to accommodate the computing resource requirements of different scenarios, we also provide an alternative v2 version for a trade-off between computation complexity and sound quality. In decoder-v2, we only use iSTFT-based decoder-v1 to model the high-frequency part while use the upsampling layer with a residual network in the GAN-based vocoder to model the low-frequency part. Because the upsampling method can synthesize high-quality harmonics and the high-frequency part requires less modeling capability, iSTFT is sufficient to meet its modeling requirements. Then, we take the signals generated by the upsampling network as low-frequency bands, and the signals generated by the decoder-v1 part as high-frequency bands, and adopt the pseudo quadrature Mirror filter-bank (PQMF) for subband modeling.
\vspace{-3pt}
\vspace{-0.3cm}

\vspace{-12pt}
\subsection{Discriminator}
\label{sec2:discriminator}
A multi-resolution spectrum discriminator (MSD)~\cite{DBLP:journals/corr/abs-2011-09631} and a multi-resolution complex-valued spectrum discriminator (MCD) are adopted for the adversarial training. The models frequency domain information at different levels, and the discriminator is particularly effective in the iSTFT decoder and has noticeable gains in high-frequency harmonic reconstruction. In addition to MSD, the proposed MCD is to model the relationship between the real and imaginary parts of the signal, which is useful for improving the phase accuracy. To be specific, MCD divides the signal into real and imaginary parts by short-time Fourier transform (STFT) at multiple scales and then applies 2-d complex convolutions to the input, which has been shown to work well in the complex-valued domain~\cite{DBLP:conf/interspeech/HuLLXZFWZX20}. Its architecture is consistent with the multi-resolution spectrum discriminator.

\vspace{-3pt}
\subsection{Loss}
\label{sec2:Loss}

The training of AdaVITS includes CVAE and GAN training. The CVAE loss is expressed as
\begin{equation}
  \setlength{\abovedisplayskip}{3pt}
  \setlength{\belowdisplayskip}{3pt}
\begin{split}
    L_{\text{cvae}} = L_{\text{kl}}+\lambda_{\text{recon}} * L_{\text{recon}} +\lambda_{\text{ppg}} * L_{\text{ppg}}
\end{split}
\end{equation}
where $L_{\text{kl}}$ is Kullback-Leibler divergence following VITS, and $L_{\text{recon}}$ calculates the L1 distance between the mel-spectrum of the waveform generated by the decoder and the ground truth. ${\lambda}_{\text{recon}}$ and ${\lambda}_{\text{PPG}}$ are 45 and 10, respectively. With GAN training, the final objectives are expressed as
\begin{equation}
  \setlength{\abovedisplayskip}{3pt}
  \setlength{\belowdisplayskip}{3pt}
\begin{split}
    L_{\text{G}}= L_{\text{adv}}(\text{G}) + \lambda_{\text{fm}} * L_{\text{fm}}(\text{G}) + L_{\text{cvae}}
\end{split}
\end{equation}
\begin{equation}
  \setlength{\abovedisplayskip}{3pt}
  \setlength{\belowdisplayskip}{3pt}
\begin{split}
    L_{\text{D}} = L_{\text{adv}}(\text{D})
\end{split}
\end{equation}
where $L_{\text{adv}}(\text{G})$ and $L_{\text{adv}}(\text{D})$ are the GAN loss of G and D, the feature matching loss $L_{\text{fm}}$ is used to improve the training stability, and the ${\lambda}_{\text{fm}}$ is 2. Since the discriminator consists of multiple sub-discriminators in the MSD and MCD, the above GAN loss and feature matching loss are the sum of the losses of multiple sub-discriminators.

\vspace{-5pt}
\section{Experiments}
\label{sec:majhead}

\begin{table*}[t]
	\centering
	\setlength{\tabcolsep}{4.2mm}
	\caption{Experimental results in terms of MOS and WER. Model parameters and computational complexity are also shown.}
	\begin{tabular}{lclclclclclcl} 
		\toprule
		Model&Params (M) &Com. (GFlops)&Naturalness& Similarity&WER (\%) &  \\ 
		\midrule
		Fs2-o+HiFiGAN v1&40.16&15.85&3.08~($\pm$0.13)&3.21~($\pm$0.10)&8.90  \\
		FS2-l+HiFiGAN v2&8.67&0.98&2.63~($\pm$0.11)&3.08~($\pm$0.14)&10.53  \\
		VITS~\cite{DBLP:conf/icml/KimKS21}&29.36&15.76&3.59~($\pm$0.13)&3.53~($\pm$0.12)&15.29 \\
		AdaVITS-e &8.70&0.66&2.82~($\pm$0.15)&3.16~($\pm$0.16)&11.11 \\
		AdaVITS-v1 &8.97&0.72&2.94~($\pm$0.14)&3.10~($\pm$0.14)&8.19  \\
		AdaVITS-v2 &11.55&3.63&3.15($\pm$0.13)&3.12~($\pm$0.12)&8.17  \\
		\midrule
		Recording &-&-&3.70~($\pm$0.12)&3.62~($\pm$0.11)&4.68  \\
		\bottomrule
	\end{tabular}
	\label{MOS}
\end{table*}

\subsection{Datasets}
\label{sec3:Datasets}
\vspace{-5pt}

We use $train{-}clean360$ and $train{-}clean100$ subsets of LibriTTS~\cite{DBLP:conf/interspeech/ZenDCZWJCW19} for pre-training model, which contains around 242 hours of utterances from 1151 speakers. To evaluate the performance of AdaVITS in the speaker adaption task, VCTK~\cite{veaux2017superseded}, which is another commonly used multi-speaker TTS corpus with different acoustic conditions from LibriTTS, is adopted to fine-tune the pre-trained model. In practice, five males and five females are randomly selected from VCTK to work as the target speakers for speaker adaptation. For each speaker, 20 utterances are randomly selected. Another 10 extra sentences from each speaker are randomly selected, resulting in a testing set with a total of 100 sentences from 10 speakers.

All audio samples are downsampled to 16kHz, and are then represented as frame level with 12.5ms hop length and 50ms window length.  We use the bottleneck feature extracted from the pre-trained WeNet~\cite{DBLP:conf/interspeech/YaoWWZYYPCXL21} model as PPG with dimension 256.

\vspace{-5pt}
\subsection{Model Configuration}
\label{sec3:Model Configuration}
\vspace{-3pt}

To evaluate the performance of AdaVITS, some representative models, including \textit{Fastspeech 2 with HiFiGAN} and \textit{VITS} are compared in the experiments. In practice, two \textit{Fastspeech 2 with HiFiGAN} systems are compared, which are referred to as \textit{Fs2-o+HiFiGAN v1} and \textit{FS2-l+HiFiGAN v2}, respectively. Fs2-o is a standard Fastspeech 2 model, which follows the basic architecture in Fastspeech 2~\cite{DBLP:conf/iclr/0006H0QZZL21}. The difference is that the duration predictor and pitch predictor use 5 Conv1D layers with kernel size 5 for more prediction accuracy. Compared with the original Fastspeech 2, the energy predictor is not used in Fs2-o. Fs2-l is a lightweight version with reduced filter size and layers. Compared with Fs2-o, we use two FFT layers in both encoder and decoder of Fs2-l and set the filter size and hidden dimension to 768 and 128. As for the vocoder, compared with HiFiGANv1, HiFiGANv2 is a lightweight version, and the details can be found from ~\cite{DBLP:conf/interspeech/SuJF20}. This Fs2-l+HifiGANv2 has similar model parameters and computational complexity with the proposed AdaVITS, designed for comparison purpose.



In the AdaVITS, AdaVITS-v1 means to use decoder-v1, AdaVITS-v2 means to use decoder-v2 described in Section 2.2.2. The duration predictor of AdaVITS and all FFT blocks follow the configuration of Fs2-l. All encoders in the proposed approach consists of 2 FFT blocks and the post-net follows~\cite{DBLP:conf/nips/RenRTQZZL19}. In the decoder, for the decoder-v1, conv1D channels are [256, 384, 1026], and all kernels are set to 3; and for the decoder-v2 , conv1D channels are [256, 384, 774] while the upsample rates are [5, 5, 2] and upsample hidden channels are [256, 192, 64]. Through this method, decoder-v2 upsampling layers model low frequency from 0 to 4Khz, and iSTFT modeling high frequency from 4 to 16Khz. We follow \cite{yang2021multi} for all upsampling layers and residual network structures. The MCD and MSD follow the architecture of MSD in Glow-WaveGAN\cite{DBLP:conf/interspeech/CongYXS21}. Other settings are the same as the original VITS\cite{DBLP:conf/icml/KimKS21}. 

In addition to AdaVITS, a variation referred to as AdaVITS-e is also compared. In AdaVITS-e, the model is trained via an end-to-end way with text as input instead of using PPG as an intermediate constraint. The architecture of AdaVITS-e is similar to VISinger~\cite{DBLP:journals/corr/abs-2110-08813} but without the F0 predictor. The number of FFT block layer of the text encoder and frame prior network is set to 2, and other configurations in AdaVITS are applied to AdvaTTS-e. 

In all the above models, the dimension of speaker embedding is set as 256. Fs2-o/Fs2-l/VITS/AdaVITS-e/AdaVITS pre-trained models and HiFiGAN v1/HiFiGAN v2 are trained up to 800k steps on 2080Ti GPU with batch size of 32. In the adaptation process, we finetune Fs2-o/Fs2-l/VITS/AdaVITS/AdaVITS-e on 2080Ti GPU for 2000 steps, and HiFiGAN v1/HiFiGAN v2 will not be updated further.

\vspace{-5pt}
\subsection{Experimental Results}
\label{sec3:Evaluate}
To evaluate the performance of different models, a mean opinion score (MOS) test is conducted in terms of the naturalness and speaker similarity. A good synthesized sample should have high quality in naturalness and similarity with the target speaker. In this human rating test, each utterance is listened by 20 listeners, and the participants are asked to rate the sample with a score ranging from one to five for the naturalness and speaker similarity respectively.  As for the evaluation of computational complexity, the GFlops required 
for generating speech per second is utilized. Since the computational complexity required by scaled-dot attention is not linear with the sentence length, the average of the test set is used as the result. In addition, the word error rate (WER) of each system is calculated to show the stability of each model, especially concerning pronunciation and intonation. WER is calculated by pre-trained WeNet model\footnote{\url{https://github.com/wenet-e2e/wenet/tree/main/examples/librispeech/}}. Note that 
this model is different from the model used to extract PPG.  Results are shown in Table~\ref{MOS}.

As can be seen from Table~\ref{MOS}, compared with \textit{FS2-l+HiFiGAN v2} which has a similar model size with \textit{AdaVITS-v1}, the proposed AdaVITS achieves better naturalness and less computational complexity. As for the WER of samples synthesized by AdaVITS is only 52.6\% of that synthesized by \textit{FS2-l+HiFiGAN v2}, indicating the good stability of AdvaVITS. Compared with \textit{FS2-o+HiFiGAN v1}, \textit{AdaVITS-v2} has a similar naturalness but a smaller model size. Compared with the original VITS, AdaVITS still has a gap to bridge in terms of naturalness and speaker similarity. However, AdaVITS achieves much better WER compared with other methods, which is mainly attributed to the utilization of PPG-based linguistic features, which can be proved by the performance of AdaVITS-e, in which regular text is used as input. It should be noted that the higher MOS score in terms of naturalness has no necessary relationship with the WER, as the participants paid more attention to the prosody and quality of speech and our human beings have a higher tolerance for the pronunciation than an ASR model.  

\vspace{-5pt}
\subsection{Ablation study}
\label{sec3:Ablation study}
To evaluate the effectiveness of each component in the proposed method, an ablation study is conducted by dropping out each component respectively on AdaVITS-v1. To be specific,  the effectiveness of linear attention, FLE, MCD,  PPG predictor, and iSTFT decoder is analyzed, and the results can be found in Table~\ref{MOS for ablation study}. As can be seen, MCD and PPG predictor play important roles in obtaining high quality speech, while the addition of FLE, linear attention, and iSTFT decoder can effectively reduce the number of parameter or computation complexity without leading to a obvious impact on the results.

\begin{table}[h]
    
	\centering
	\caption{Ablation study results. The MOS test are for the naturalness. w/o means without. In w/o~Linear Att., the linear attention is replaced by the scaled-dot attention. In w/o~ISTFT Dec, the ISTFT decoder is replaced by the decoder of HiFiGAN v2.}
	\setlength{\tabcolsep}{3mm}{
	\begin{tabular}{lclclcl}
		\toprule
		Model & Params  & Com.  &MOS  \\ 
		\midrule
		
		AdaVITS-v1 &8.97 & 0.72&3.17~($\pm$0.12)  \\
        \ w/o~Linear Att.  &8.97 & 0.83  &3.14~($\pm$0.13)  \\
        \ w/o~FLE         &11.88 & 0.72  &3.18~($\pm$0.14)  \\
        \ w/o~MCD &8.97 & 0.72  &2.99~($\pm$0.14)  \\
        \ w/o~PPG Predictor &8.97 & 0.72  &2.83~($\pm$0.13)  \\
        \ w/o~ISTFT Dec. &7.24 & 1.46  &3.32~($\pm$0.13) \\
        \midrule
        Recording &-&-&4.04~($\pm$0.10)  \\
		\bottomrule
	\end{tabular}
	}
	\label{MOS for ablation study}
\end{table}
\vspace{-0.5cm}

\section{Conclusions}
\label{sec:Conclusions}
In this paper, a VITS-based lightweight adaptive TTS system, referred to as AdaVITS, is proposed to support speaker adaption's need for low cost. To effectively reduce the computational complexity leaded by the upsampling-based vocoder, an iSTFT-based wave construction decoder is proposed. In addition, NanoFlow is utilized to reduce the parameters of the prior encoder, and scaled-dot attention in FFT is replaced with linear attention to further reduce the computational complexity. To ensure the stability of the simplified model, PPG is used as frame-level linguistic features. Extensive experiments demonstrate the obvious superiority of AdaVITS in terms of the model size and computational complexity compared with other standard models. When compared to the model with similar parameters, the proposed AdaVITS achieves less computational complexity and better speech quality.

\bibliographystyle{IEEEtran}

\bibliography{mybib}


\end{document}